\documentstyle[prl,aps,preprint,epsfig]{revtex}

\begin{document}



\title{Diffusion constant for the repton model of gel electrophoresis}
\author{M. E. J. Newman}
\address{Santa Fe Institute, 1399 Hyde Park Road, Santa Fe, NM 87501.
U.S.A.}
\author{G. T. Barkema}
\address{HLRZ, Forschungszentrum J\"ulich, 52425 J\"ulich, Germany}
\date{March 6, 1997}
\maketitle

\begin{abstract}
  The repton model is a simple model of the ``reptation'' motion by which
  DNA diffuses through a gel during electrophoresis.  In this paper we show
  that the model can be mapped onto a system consisting of two types of
  particles with hard-sphere interactions diffusing on a one-dimensional
  lattice.  Using this mapping we formulate an efficient Monte Carlo
  algorithm for the model which allows us to simulate systems more than
  twice the size of those studied before.  Our results confirm scaling
  hypotheses which have previously been put forward for the model.  We also
  show how the particle version of the model can be used to construct a
  transfer matrix which allows us to solve exactly for the diffusion
  constant of small repton systems.  We give results for systems of up to
  20 reptons.
\end{abstract}



\newpage

\section{Introduction}
\label{intro}
Gel electrophoresis is an experimental tool of great importance in the
ever-growing industries of genetics and polymer research.  It allows an
experimenter to separate a mixture of polymer strands by length or, in
combination with sample strands of known lengths, to measure the lengths of
strands in the mixture.  However, despite the development of sophisticated
techniques for exploiting electrophoresis in the laboratory, the physics
behind the method is still imperfectly understood.  A number of efforts
have been made to model the diffusion of polymers in gels, with varying
degrees of success.  In this paper we examine one of the simplest but also
most successful of these, the repton model, which is primarily of use as a
model of DNA electrophoresis in agarose gels.  (It does not model the
behavior of most other polymers very well, nor the behavior of DNA in
polyacrylomide gels.)

Although the subject of many functional refinements and improvements over
the years, the basic DNA/agarose gel experiment is in essence a very
simple one.  A gel is formed by adding agarose powder to a warm buffer
solution, pouring the mixture into the gel box, and allowing it to cool,
whereupon the agarose strands cross-link to form a three-dimensional web
with a typical pore size of about 1000~\AA.  The DNA is placed in solution
with more of the same buffer, from which it gains an electric charge, and
is then injected into the gel.  An electric field, typically on the order
of a few volts per centimeter, is applied horizontally across the box, and
the charged DNA migrates under its influence through the gel.  It is found
that the rate of migration depends principally on three factors: the pore
size of the gel (which in turn depends on the concentration of the
original agarose solution), the magnitude of the applied electric field,
and, crucially, the length of the DNA strands.  Longer DNA strands are
impeded more by the agarose in the gel and travel slower, so that after
the experiment has been running for some time (typically a few hours), the
initially homogeneous DNA mixture will have separated out along a ``lane''
in the gel box, according to the lengths of its constituent parts.  The
distance traveled by any particular component of the mixture is a measure
of its length, and the final gel can be dissected into pieces to refine
the DNA by length.

The basic mechanism responsible for the length-dependence of the migration
rate was first described by de Gennes\cite{deGennes71}.  The important
point is that the persistence length of DNA under the conditions of the
experiment is between about 400 and 800~\AA, which is of the same order of
magnitude as the pore size of the gel.  This means that the agarose strands
in the gel effectively prevent movement of a strand of DNA transversely to
its length, so that it migrates primarily by a longitudinal slithering
motion, for which de Gennes coined the term ``reptation''.  The repton
model is one of the simplest models of the reptation process.  It was
invented in 1987 by Rubinstein\cite{Rubinstein87} as a model of the
dynamics of block co-polymers, and co-opted as a model of DNA reptation by
Duke\cite{Duke89}.  Briefly the model is as follows.  (A more thorough
exposition can be found, for example, in Ref.~\onlinecite{BN97}.)

The repton model simulates the movement of a single strand of DNA through
the gel.  The strand is represented as a number $N$ of points, joined
together by lines---see Figure~\ref{reptonmodel}(a).  The points are
confined to the squares of a two-dimensional rectangular lattice which is
represented by the grid of lines in the figure.  The points in the strand
are known as ``reptons'', and successive reptons in the strand may lie
either in the same square, or in adjacent ones, but may not be further
apart than this.  This ensures that the DNA has some elasticity but is not
infinitely stretchy.  The lattice squares represent the pores in the
agarose gel, so that the lattice parameter $a$ is also the pore size.  The
distance between successive reptons in the chain represents the persistence
length of the DNA (a few hundred base pairs, or about 400 to 800~\AA, as
mentioned above), and since this is limited to at most one lattice spacing,
it is implicit in the model that the persistence length and the pore size
are approximately equal.  (It is for this reason that the repton model
makes a poor approximation to the behavior of DNA in other types of gel, or
of other polymers in agarose.)

The use of a two-dimensional lattice in the model appears at first to be
problematic.  Clearly the real DNA moves in a three-dimensional space.
However, as we will see in Section~\ref{mappings}, the behavior of the
model is actually independent of the dimensionality of the lattice, so a
two-dimensional one is as good as any other.

In order to simulate the reptation motion of the DNA, the following
dynamics is imposed on the repton chain.
\begin{enumerate}
\item A repton in the interior of the chain may move to one of the
adjacent squares on the lattice, provided that one of its immediate
neighbors in the chain is already in the square to which it is moving, and
the other is in the square which it leaves.
\item The two reptons at the ends of the chain may move in any direction
to an adjacent square, provided that such a move does not take them more
than one lattice spacing away from their neighboring repton on the chain.
\end{enumerate}
These two rules ensure that the chain always diffuses along its own length,
and that neighboring reptons on the chain are never more than one lattice
spacing apart.  The time-scale for the model is set by stipulating that
each repton, driven by random thermal fluctuations, should attempt to move
in each of the four possible directions once on average per unit time.
Each attempted move is accepted and carried out if it is allowed by the
rules above, otherwise it is rejected and the chain stays as it was.

Notice that all allowed moves proceed at the same rate; there are no
energies in the repton model and no Boltzmann weights.  The model is
entirely entropy-driven.  If we wish to model the migration of the DNA in
the applied electric field we need to introduce a bias in the selection
probabilities for moves in one particular direction, usually the $x$ axis.
However, if, as the experimentalists are, we are primarily interested in
the rate $v$ of migration in a given electric field $E$, it is not
necessary to introduce such a bias.  The value of $v$ in the low-field
limit can be calculated instead from the zero-field diffusion constant $D$
using the Nernst-Einstein relation:
\begin{equation}
D = \lim_{E\to0} {v\over NE}.
\label{nerel}
\end{equation}
This is the approach we will be taking in this paper, and our calculations
will concentrate on the evaluation of the diffusion constant, in the
knowledge that the value of $v$ can always be calculated from it in a
simple fashion.

The repton model ignores many important features of the dynamics of the
real DNA, such as excluded volume effects due to the finite space occupied
by the DNA, self-repulsion effects due to its charge, the effects of
counterions, mechanical properties of the DNA, and inhomogeneities in the
gel.  Surprisingly however, it gives results in fairly good agreement with
experiment\cite{BCM96}, leading us to believe that it may well capture many
of the essential features of the dynamics of DNA in agarose.

Some of the earliest analytic work on the repton model was performed by van
Leeuwen and Kooiman\cite{VK91}, who showed that under the assumption of
periodic boundary conditions the limiting value of $D$ as $N\to\infty$ is
$1/(3N^2)$.  Later, Pr\"ahofer\cite{Prahofer94} demonstrated rigorously
that this limiting behavior holds also for the repton model without
periodic boundary conditions.  In practice however, the large finite-size
effects present in the model mean that the measured value of $D$ is far
from this limit.  In our calculations we usually quote figures for $DN^2$,
which should have a limiting value of $\frac13$.

In 1991, Widom~{\it{}et~al.}\cite{WVD91} took an important step forward by
showing that it was possible to solve the repton model exactly for finite
values of the chain length $N$ using a combination of a transfer matrix
method with a perturbation theoretic analysis.  The matrix used is
essentially the stochastic or Markov matrix for the dynamics of the model,
which has one row and one column for each possible state of the chain.  As
discussed in Section~\ref{analytics}, this matrix, in its most compact
form, has rank $3^{N-1}$.  Unfortunately this makes the required
diagonalization operation prohibitively costly for all but the shortest
repton chains.  Widom~{\it{}et~al.}\ carried out the calculation for values
of $N$ up to 5, and these results were later extended by Szleifer and
Bisseling to $N=12$ using a special-purpose computer\cite{SB91}.

In this paper we propose a new mapping of the repton model onto a
one-dimensional particle model with two types of particles possessing
hard-sphere repulsion.  Using this mapping we show that the repton model
can be solved exactly by finding only the one eigenvector corresponding to
the largest eigenvalue of a much smaller matrix, one with rank $(N+1)
2^{N-2}$.  Given the relatively conservative size of this matrix and the
fact that finding one eigenvector is a lot simpler than diagonalizing the
whole matrix, we have been able using this method to extend the exact
solution of the model to chains of up to twenty reptons (corresponding to
DNA strands of about 4~kb), using only conventional computing resources.

Of the numerical studies which have been performed on the model, probably
the most comprehensive to date are those of Barkema, Marko, and
Widom\cite{BMW94}, who used a multi-spin coded algorithm running on a
supercomputer to simulate the model for values of $N$ up to 100.  Our
projection of the model onto a one-dimensional particle system has
also led us to a more efficient Monte Carlo algorithm for the model's
simulation which has allowed us to extend these simulation results to
$N=250$, again using only conventional computing resources.

The outline of this paper is as follows.  In Section~\ref{mappings}, we
describe the mapping introduced by Duke\cite{Duke89} of the repton model
onto a one-dimensional chain model, and then introduce our further mapping
onto a model containing two types of particles diffusing on a
one-dimensional lattice.  In Section~\ref{numerics} we describe our
simulations of this version of the model.  In Section~\ref{analytics} we
make use of the mapping to define our reduced transfer matrix and from that
matrix extract exact results for the diffusion constant of the model for
values of $N$ up to 20.  We also speculate on the connection between the
particle version of the repton model and the so-called ``asymmetric
exclusion models''.  Some simple versions of these models have been solved
analytically for all $N$\cite{DEHP93,SS95}, which leads us to hope that a
similar solution of the repton model may be possible.  In
Section~\ref{concs} we give our conclusions.

\section{The projected repton model}
\label{mappings}
\begin{figure}
\begin{center}
\psfig{figure=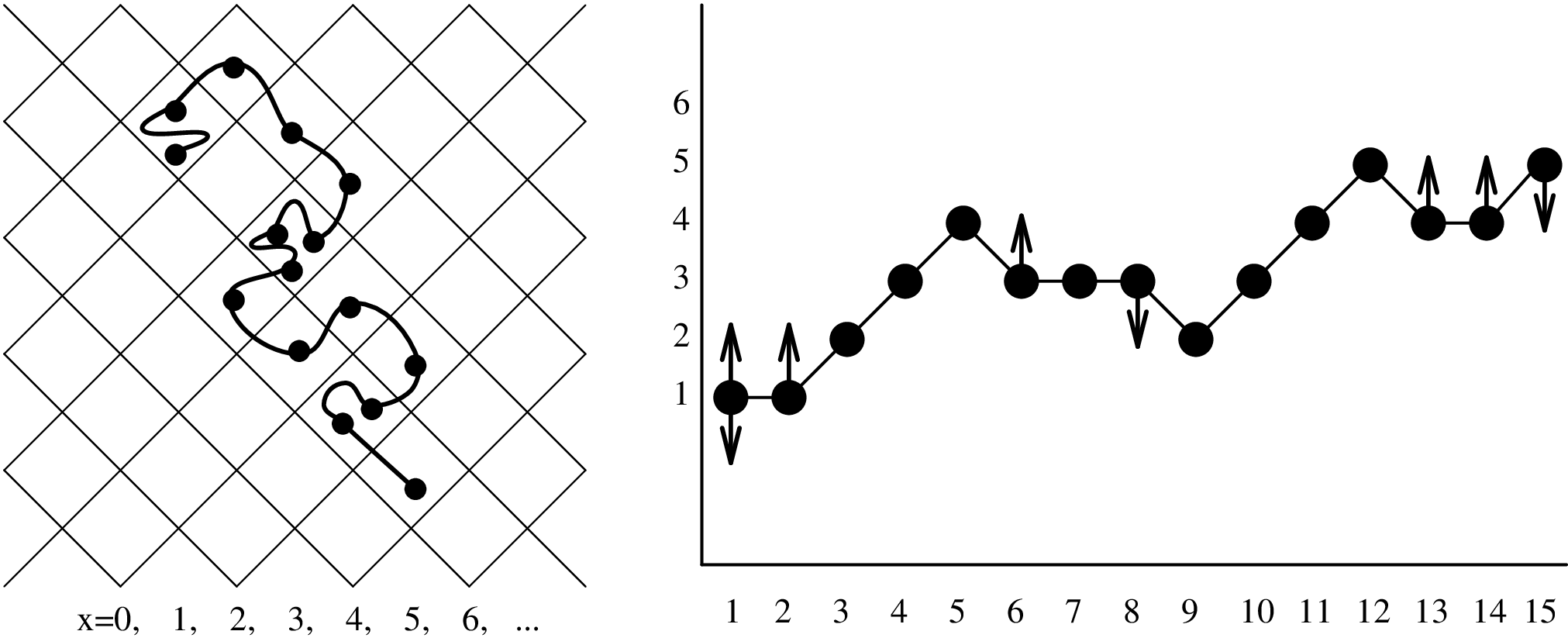,width=\textwidth}
\end{center}
\caption{(a) The repton model of gel electrophoresis.  The
  points---reptons---are connected together in a chain, and consecutive
  points in the chain must occupy adjacent squares on the lattice, or the
  same square.  (b) The projected repton model described in
  Section~\ref{mappings}.  The arrows indicate the possible moves.
\label{reptonmodel}}
\end{figure}

The fundamental quantity which we would like to calculate using the repton
model is the rate of migration of DNA, as a function of its length, under
the influence of an electric field $E$ applied, for example, horizontally
in Figure~\ref{reptonmodel}(a).  As we pointed out in the last section,
this can be calculated using the Nernst-Einstein relation,
Equation~(\ref{nerel}), from a knowledge of the diffusion constant for the
diffusion of the repton chain in zero field.  Since we are only interested
in the movement of the chain along one axis, in this case the $x$ axis, it
is only necessary to consider the $x$ component of each repton's position.
In Figure~\ref{reptonmodel}(b) we have plotted this $x$ component for the
state of the chain depicted in Figure~\ref{reptonmodel}(a) as a function of
position along the chain from one end to the other, and this plot contains
all the information we need about that state.  This projected form of the
repton model was first introduced by Duke~\cite{Duke89}, and it is this
form which was used by Widom~{\it{}et~al.}\ to construct their transfer
matrix, and also by Barkema~{\it{}et~al.}\ to perform simulations of the
model's properties.

The restriction that consecutive reptons in the chain can lie only in the
same or adjacent squares on the lattice translates to the restriction that
the positions $x$ of adjacent reptons in the projected model can differ by
at most $a/\sqrt{2}$, although we will find it more convenient to measure
$x$ in units of this quantity, so that the values are always integers and
adjacent ones may differ by only $+1$, $0$, or $-1$.  The dynamics of the
projected model in zero electric field also can be derived as a simple
projection of the dynamics of the original model:
\begin{enumerate}
\item A repton in the interior of the chain may move up or down by one
step provided that one of its two neighbors is already at the level to
which it is moving, and the other is at the level which it leaves.
\item The two reptons at the ends of the chain may move either up or
down, provided this does not take them more than one step away from their
neighbors.
\end{enumerate}
As with the two-dimensional version of the model, each possible move of
each repton is attempted on average once per unit time, and any attempted
move is accepted provided it doesn't violate any of the rules above.  If
it does, the move is rejected and the chain remains unchanged.

Notice that this projection of the repton model onto the $x$ axis would
work just as well had we started with a lattice with three or even more
dimensions.  We conclude that the dynamics of the model is independent of
the dimensionality of the original lattice, as we mentioned briefly in the
last section.

\begin{figure}
\begin{center}
\psfig{figure=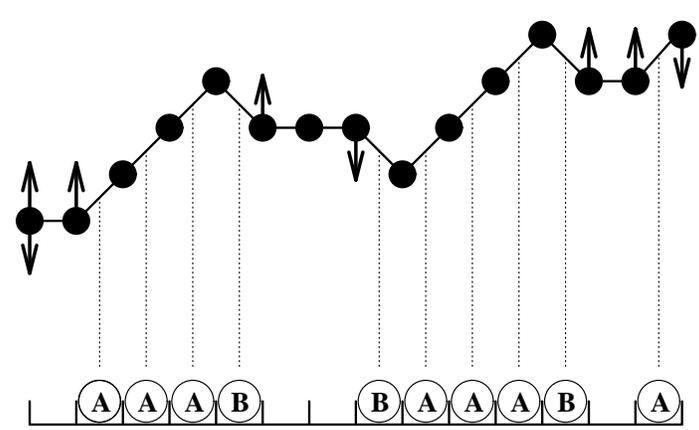,width=9cm}
\end{center}
\caption{The mapping of the projected repton model onto a particle model.
  There are two types of particle, which we label A and B.  No two
  particles may coexist on the same lattice site, but otherwise they are
  free to diffuse about the lattice.
\label{particles}}
\end{figure}

The state of the projected repton model can be specified by giving the $x$
coordinate of each of the $N$ reptons in the chain.  Alternatively, as
Figure~\ref{reptonmodel}(b) makes clear, we could specify it by describing
in turn the $N-1$ links in the chain---the lines between adjacent reptons
in the figure---each of which can be in one of three states: sloping
upwards to the left, sloping upwards to the right, or level.  In order to
completely pin the chain down, we would also have to specify the absolute
position of one of the reptons---say the left-most one---but since the
model is translationally invariant, all its properties can be calculated
without knowledge of this variable.  In Figure~\ref{particles} we have
made use of these new degrees of freedom to create an alternative mapping
of the model to a one-dimensional particle model.  In this mapping each of
the $N-1$ links corresponds to a site on a new lattice, and each site can
be in one of three states of occupation, depending on the state of the
corresponding link.  Sites corresponding to links which slope upwards to
the left are occupied by a particle of one type, which we call type~A, and
those corresponding to links which slope upwards to the right are occupied
by a different type of particle, type~B.  Sites corresponding to
horizontal links are left empty.  It is not hard to show that the dynamics
of the particles is as follows:
\begin{enumerate}
\item Particles in the interior of the chain are conserved.
\item No two particles may coexist at the same site, regardless of their
types.
\item A particle adjacent to an empty site can move to that site.
\item A particle at one of the ends of the chain can fall off the chain
and vanish.
\item If one of the end sites on the chain is empty, a new particle of
either type can appear there.
\end{enumerate}
As before, all possible moves are attempted once each on average per unit
time.  Moves which violate none of the rules above are always accepted.
All others are rejected.  The particles are in some ways akin to Fermions,
but their dynamics is completely classical in nature, so we prefer to
regard them simply as hard-sphere classical particles.

As far as the diffusion of the repton chain is concerned, we can show that
if a particle of type~A enters the system at the left-hand end and
migrates all the way to the right, and falls off, then the average
position of the chain moves one step in the positive $x$ direction.  It
also moves one step in the positive $x$ direction if a particle of type~B
moves across the system from right to left.  Moves in the opposite
directions correspond to motion of the average position in the negative
$x$ direction.

By employing this particle mapping of the repton model, we have been able
to improve considerably on previous calculations of the model, both
numerical and exact.

\section{Numerical calculations}
\label{numerics}
If we were to simulate our particle version of the repton model directly,
employing the dynamics as described above, the resulting calculation would
be entirely equivalent to, and no faster than, a direct simulation of the
projected repton model.  However, we can speed the calculation considerably
by observing that, in the zero-field case we are considering here, the
dynamics of the two types of particles, A~and~B, is identical.  Whether a
particular particle is of type~A or of type~B makes no difference
whatsoever to the probability of the system taking a particular path.
Thus, it is possible to carry out the entire simulation without assigning
any types to any of the particles---we can assign types to them at the end
instead.  In fact, the best statistics are derived by making in turn every
possible assignment of particle types to particles and averaging over all
of them.  If we denote by $n_{A\leftarrow}$ and $n_{A\rightarrow}$ the
number of particles of type~A which pass through the system from left to
right and from right to left respectively during the course of our
simulation, and similarly for particles of type~B, then the mean square
distance $\langle d^2 \rangle$ traveled by the repton chain in the $x$
direction, where $\langle\ldots\rangle$ indicates an average over all
possible assignments of particle types, is
\begin{eqnarray}
\langle d^2 \rangle &=& \langle [(n_{A\rightarrow} + n_{B\leftarrow}) -
                        (n_{A\leftarrow} +
                        n_{B\rightarrow})]^2 \rangle\nonumber\\
                    &=& \langle [(n_{A\rightarrow} - n_{B\rightarrow}) -
                        (n_{A\leftarrow} -
                        n_{B\leftarrow})]^2 \rangle\nonumber\\
                    &=& \langle (n_{A\rightarrow} -
                        n_{B\rightarrow})^2 \rangle +
                        \langle (n_{A\leftarrow} -
                        n_{B\leftarrow})^2 \rangle\nonumber\\
                    &=& \langle (n_{A\rightarrow} +
                        n_{B\rightarrow}) \rangle +
                        \langle (n_{A\leftarrow} +
                        n_{B\leftarrow}) \rangle\nonumber\\
                    &=& n_\rightarrow + n_\leftarrow.
\label{dsqr}
\end{eqnarray}
The third line here follows from the statistical independence of the
numbers of particles of each type, and the fourth line follows from the
properties of random walks in one dimension.  The variables $n_\rightarrow$
and $n_\leftarrow$ in the last line are the total numbers of particles
passing across the system in each direction during the course of the
simulation.  The diffusion constant in the $x$ direction is related to
$\langle d^2 \rangle$ by
\begin{equation}
D = {\langle d^2 \rangle\over2 t}
\label{diff}
\end{equation}
where $t$ is the length of time (as defined in Section~\ref{intro}) for
which the simulation ran.

In order to calculate $n_\rightarrow$ and $n_\leftarrow$, we need to count
the number of particles falling off the right-hand end of the system which
entered from the left, and the number falling off the left which entered
from the right.  Those which enter and leave at the same end make no
contribution.  Thus we need to keep a record for each particle in the
system of whether it entered at the left- or the right-hand end.  We can do
this by labeling each with either an ``L'' or an ``R'', which in effect
means that we have two types of particles again.

\begin{figure}
\begin{center}
\psfig{figure=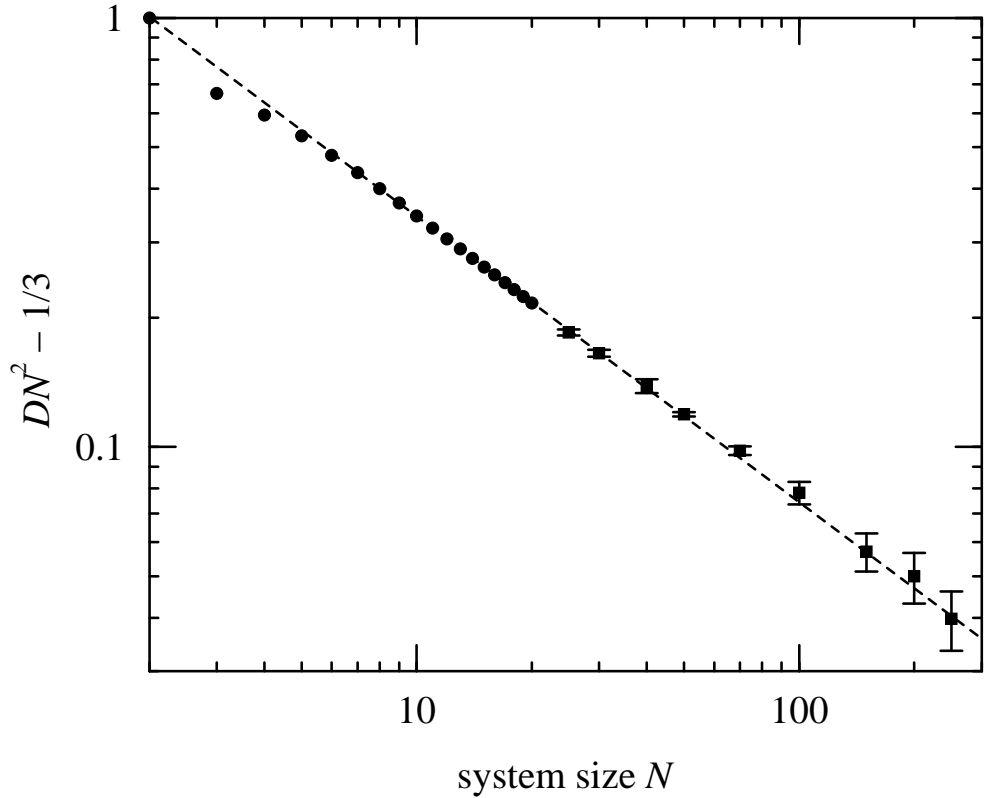,width=13cm}
\end{center}
\caption{Results for the diffusion constant $D$ of the repton model for
  systems of up to $N=250$ reptons.  The circles are exact results
  calculated using the transfer matrix method of
  Section~\protect\ref{analytics}.  The squares are numerical results
  obtained by the Monte Carlo method described in
  Section~\protect\ref{numerics}.  We have plotted the data as
  $DN^2-\frac13$ against $N$ on logarithmic scales in order to test the
  scaling form hypothesized in Ref.~\protect\onlinecite{BMW94}.  The dashed
  line indicates the slope expected if this scaling form holds.
\label{numres}}
\end{figure}

Using a multi-spin coded program which simultaneously performs 32
simulations using the algorithm described here, we have calculated the
value of the diffusion constant for the repton model for systems of up to
$N=250$ reptons.  The runs were of a variety of lengths up to a maximum of
around $10^{11}$ Monte Carlo steps for the largest systems.  For each run
we discarded the first 10\% of the data to allow for equilibration.  The
results of the simulations are shown in Figure~\ref{numres}.  Barkema,
Marko, and Widom\cite{BMW94} have conjectured, on the basis of numerical
results for systems of up to $N=100$ reptons, that $DN^2-\frac13$ scales
for large system sizes as $N^{-2/3}$.  This scaling is indicated as the
dashed line in the figure.  As the graph shows, our new results for values
of $N$ above 100 appear to confirm this conjecture.

\section{Exact calculations}
\label{analytics}
The L- and R-type particles introduced in the last section are interesting
in their own right, since, as we now show, they lead to an exact solution
of the model for small values of $N$.  Their dynamics is almost the same as
that of the A- and B-particles of the last section: they are conserved, can
hop either left or right into empty spaces, and particles at the end of the
chain can fall off altogether, and all these moves are attempted once on
average per unit time.  However, if one of the sites at the end of the
chain is empty, a new particle appears and fills it on average {\em
  twice\/} per unit time, rather than once as before.  To see this, recall
that particles of types~A and~B both attempted to fill empty end sites once
each on average per unit time in our previous version of the model.  The
total rate of attempted particle entry at either end is therefore two
particles per unit time.  Other than this, things remain as before: all
attempted moves are accepted if they don't violate any of the dynamical
rules, otherwise they are rejected.

The total rate then at which particles of type~L fall off the right-hand
end of the chain is simply one times the density of L-particles at the
rightmost site, and similarly for particles of type~R at the left-hand
end.  Since the system is, on average, completely left/right symmetric, we
can define a density function $\rho(i)$ such that
\begin{equation}
\rho_L(i) = \rho_R(N-i) = \rho(i),
\label{rhoi}
\end{equation}
where $\rho_L(i)$ and $\rho_R(i)$ are the densities of the two types of
particles at the $i$th site (counting, let us say, from the left).  Now
using Equations~(\ref{dsqr}) and~(\ref{diff}) we can write the diffusion
constant as
\begin{equation}
D = {n_\rightarrow + n_\leftarrow\over2t} = {2\rho(1) t\over2t} = \rho(1).
\label{newdiff}
\end{equation}

It would be possible to calculate $\rho(1)$, the density of the minority
particles at the first site on the lattice, using a Monte Carlo technique.
However, the results for $D$ would be less accurate than those calculated
with the methods of the last section.  On the other hand,
Equation~(\ref{newdiff}) does lend itself to exact calculations.  Notice
that the number of states of the model with the L- and R-particles is
considerably smaller that the number of states of the one with the A- and
B-particles.  The reason is that, since all particles of type~L enter from
the left, all type~R ones from the right, and the two cannot pass one
another, all lawful states of the system consist of at most two domains,
one on the left containing only L-particles and vacancies, and one on the
right containing only R-particles and vacancies.  Thus, once we know the
position of the line which separates these two domains, the state of any
particular site is completely determined if we know only whether it is
occupied or not.  There are $N$ possible positions for the line, and hence
$N 2^{N-1}$ is an upper bound on the number of states of the system.  In
fact the actual number of states turns out\cite{Note1} as we mentioned in
Section~\ref{intro}, to be $(N+1) 2^{N-2}$.  This figure is considerably
smaller than the $3^{N-1}$ states of the system with the A- and
B-particles, or equivalently of the ordinary projected repton model, which
Widom~{\it{}et~al.}\ used to construct their transfer matrix.  This
suggests that it might be possible to construct a new smaller transfer
matrix which would allow us to calculate the diffusion constant of the
model exactly for larger values of $N$.

The construction of such a matrix turns out to be charmingly
straightforward.  The matrix possesses one row and one column for each
state of the system with the L- and R-type particles.  Its off-diagonal
elements are zero except for a sparse set of elements connecting pairs of
states which are accessible to one another via the dynamics of the model.
The diagonal elements are fixed to ensure overall particle conservation.
The slowest-decaying eigenmode of this matrix represents the equilibrium
occupation probabilities of each of the states of the chain, and a simple
linear combination of its elements gives the density of particles of either
type at any site, the density $\rho(1)$ being a particular case.  We have
calculated the eigenvector of this mode numerically for systems with sizes
from $N=3$ up to $N=20$.  Since the matrix is sparse, the quickest method
of doing this is simply by repeated multiplication into an initial trial
vector.  In fact, for each system size studied, we have used two trial
vectors, chosen so that the value of $\rho(1)$ calculated from them
converges towards the equilibrium value from opposite directions, ensuring
that we have rigorous bounds on our values, as well as an absolute measure
of convergence.  The value of $\rho(1)$ can then be used in
Equation~(\ref{newdiff}) to calculate the diffusion constant.  The
resultant values for $DN^2$ are given in Table~1 and also shown in
Figure~\ref{numres}.  The values for $N$ up to 12 confirm those obtained by
Widom~{\it{}et~al.}\cite{WVD91} and by Szleifer and Bisseling\cite{SB91}.
Those for $N=13$ up to 20 are new.  No errors are given in the table; the
figures are exact to the quoted accuracy.

\begin{figure}
\begin{center}
\psfig{figure=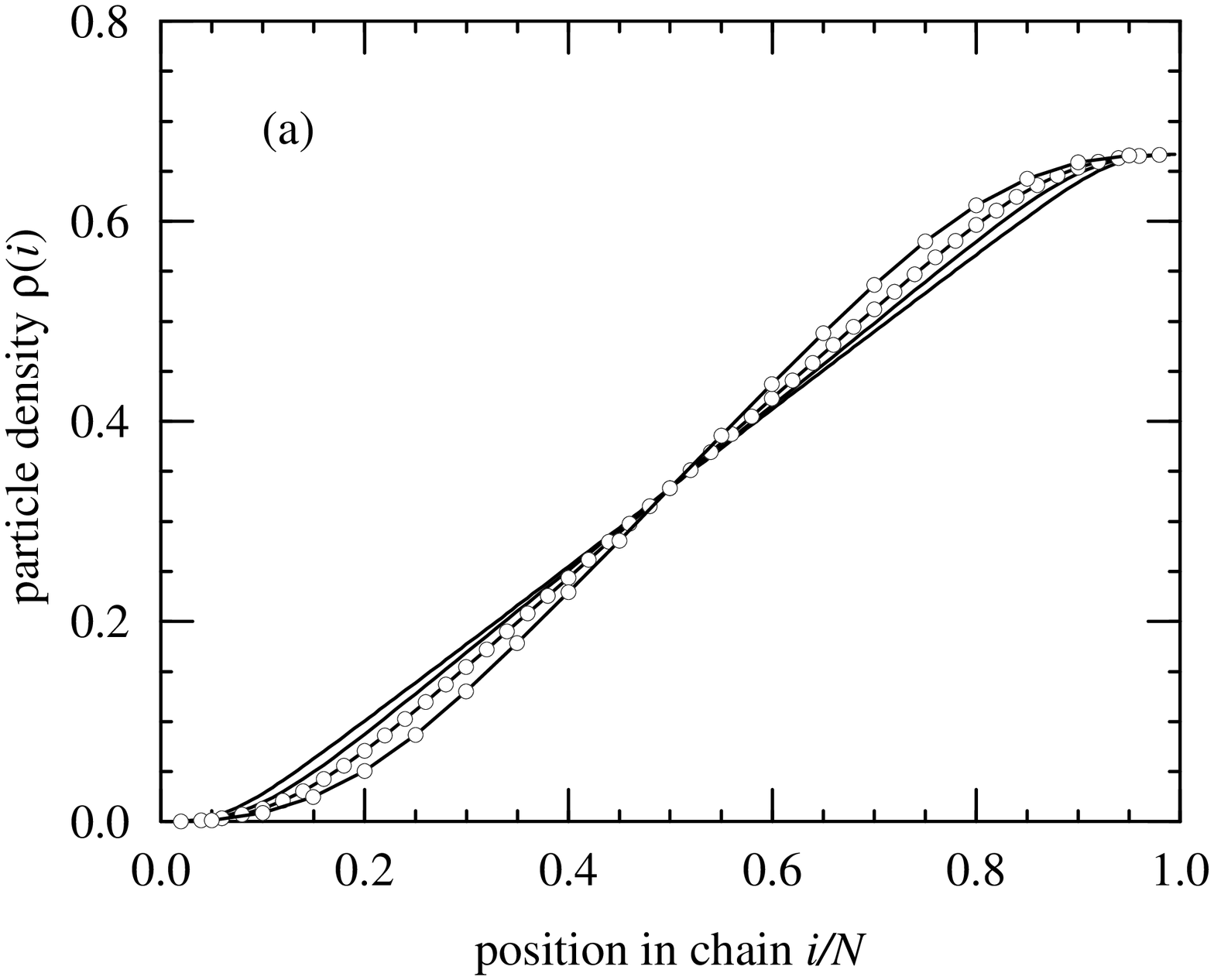,width=12cm}
\psfig{figure=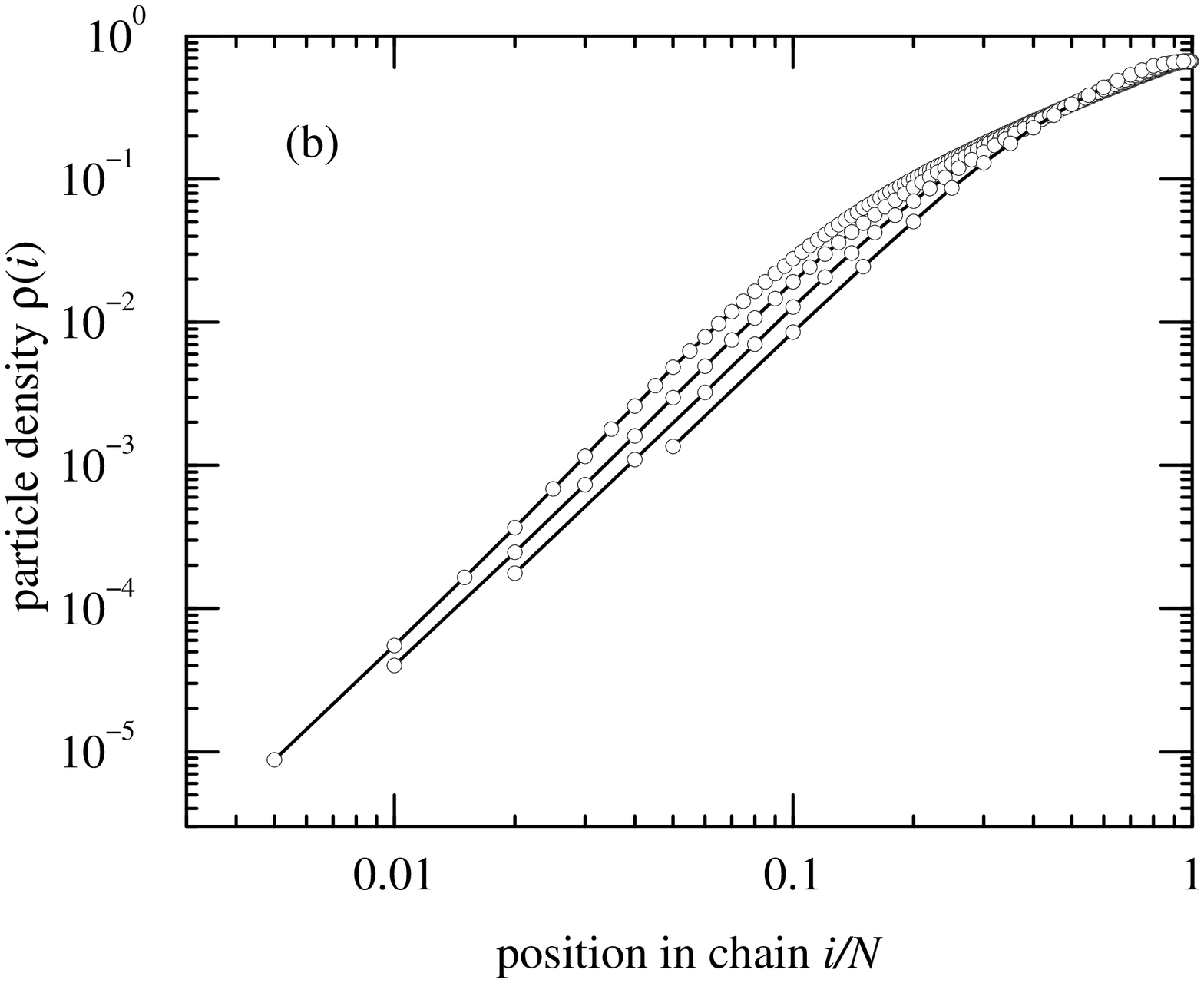,width=12cm}
\end{center}
\caption{The function $\rho(i)$ defined in Equation~(\protect\ref{rhoi}),
  plotted for systems of size $N=20$, 50, 100, and 200 on (a)~linear scales
  and (b)~logarithmic ones.  Close to the end of the repton chain (small
  $i$) $\rho(i)$ appears to vary as a power law with $i$.  The measured
  exponent of the power law is $2.76\pm0.03$.
\label{scaling}}
\end{figure}

The function $\rho(i)$ is also interesting in its own right.  In
Figure~\ref{scaling} we have plotted its value for systems of size $N=20$,
50, 100, and 200 on linear and logarithmic scales, calculated numerically.
As $N$ increases, the curves on the linear plot appear to become more and
more like a straight line with short curved regions at each end.  It seems
possible that the asymptotic form of the function is linear, with end
effects whose region of influence tends to some finite limit.  The
logarithmic plot sheds further light on the behavior at the ends of the
chain.  As it shows, $\rho(i)$ appears to become power-law in form for
small $i$ as $N$ becomes large.  The exponent of the power law is measured
to be $2.76\pm0.03$.

Consideration of $\rho(i)$ also offers a hint of a possible complete
analytic solution of the repton model.  The particle version of the model
considered here is a particular case of a class of models known as
asymmetric exclusion models.  In recent work,
Derrida~{\it{}et~al.}\cite{DEHP93} and independently Stinchcombe and
Sch\"utz\cite{SS95} have found exact analytic solutions for the density
function equivalent to our $\rho(i)$ for a number of models of this class
for all values of $N$.  The models they solved are simpler in a number of
respects than the model studied in this paper, and applying their
techniques to our model would not be a trivial task.  However,
Equation~(\ref{newdiff}) tells us that, should such a solution prove
possible, we would immediately have an expression for the diffusion
constant of the repton model for all values of $N$.  It is certainly an
intriguing prospect.

\section{Conclusions}
\label{concs}
In this paper we have studied the Duke-Rubinstein repton model of DNA
reptation in an agarose gel.  We have introduced a mapping of the model
onto a particle model in which hard-sphere particles hop at random on a
finite one-dimensional lattice.  We have demonstrated that the diffusion
constant $D$ of the polymer in the gel is proportional simply to the
average rate at which these particles cross from one side of the lattice to
the other, and we have made use of this fact to formulate a
highly-efficient Monte Carlo algorithm for calculating this diffusion
constant.  Using this algorithm we have calculated values of $D$ for
systems of up to $N=250$ persistence lengths.  These results confirm
hypotheses put forward earlier concerning the scaling of $D$ with the
system size.

We have also employed our particle version of the repton model to
construct a transfer matrix whose dominant eigenvector is directly related
to the value of the diffusion constant.  We have numerically calculated
this eigenvector, and hence extracted exact values for $D$ for systems up
to $N=20$.

Finally, we have drawn a connection between our particle version of the
repton model and the asymmetric exclusion models which have been solved
exactly by Derrida~{\it{}et~al.}\ and by Stinchcombe and Sch\"utz.  We
conjecture that it may be possible to use the techniques they employed,
along with the theory developed in this paper, to find a complete analytic
solution for the diffusion constant of the repton model for all system
sizes.

\section{Acknowledgements}
The authors would like to thank Ben Widom and Mike Widom for useful
comments and illuminating conversations.  GTB would like to thank the Santa
Fe Institute for their hospitality while this work was carried out.  This
research was funded in part by the DOE under grant number
DE--FG02--9OER40542 and by the Santa Fe Institute and DARPA under grant
number ONR N00014--95--1--0975.

After this work was completed we became aware of Ref.~\onlinecite{PS96}, in
which exact values for the diffusion constant of the repton model are
calculated for values of $N$ up to 20 using a method entirely different
from the one described here.  We thank Michael Pr\"ahofer for bringing this
to our attention.

\begin{table}
\begin{center}
\begin{tabular}{rrrrrr}
$N$ & $DN^2$ & $N$ & $DN^2$ & $N$ & $DN^2$ \\
\hline
\hline
3  & 1.000000 & 9  & 0.703951 & 15 & 0.596297 \\
4  & 0.926984 & 10 & 0.678924 & 16 & 0.585081 \\
5  & 0.864908 & 11 & 0.657549 & 17 & 0.574996 \\
6  & 0.812065 & 12 & 0.639096 & 18 & 0.565874 \\
7  & 0.769057 & 13 & 0.623006 & 19 & 0.557579 \\
8  & 0.733582 & 14 & 0.608851 & 20 & \hbox to 41pt{0.5500\hfil} \\
\end{tabular}
\end{center}
\medskip
\caption{Values of $DN^2$ calculated using the transfer matrix method
described in the text.  The values are exact to the accuracy quoted.}
\end{table}


\begin{references}
\bibitem{deGennes71}
{\frenchspacing P. G. de Gennes, J. Chem. Phys. {\bf55}, 572 (1971).}
\bibitem{Rubinstein87}
{\frenchspacing M. Rubinstein, Phys. Rev. Lett. {\bf59}, 1946 (1987).}
\bibitem{Duke89}
{\frenchspacing T. A. J. Duke, Phys. Rev. Lett. {\bf62}, 2877 (1989).}
\bibitem{BN97}
{\frenchspacing G. T. Barkema and M. E. J. Newman, Physica A, in press
(1997).}
\bibitem{BCM96}
{\frenchspacing G. T. Barkema, C. Caron and J. F. Marko,
Biopolymers {\bf 38}, 665 (1996).}
\bibitem{VK91}
{\frenchspacing
J. M. J.~van Leeuwen, J.~Phys.~I~France {\bf 1}, 1675 (1991);
J. M. J.~van~Leeuwen and A.~Kooiman, Physica A {\bf 184}, 79 (1992).}
\bibitem{Prahofer94}
{\frenchspacing M. Pr\"ahofer, Diplomarbeit, Ludwig-Maximilians-Universit\"at
M\"unchen (1994).}
\bibitem{WVD91}
{\frenchspacing B. Widom, J. L. Viovy and A. D. Defontaines, J. Phys. I
France {\bf1}, 1759 (1991).}
\bibitem{SB91}
{\frenchspacing I. Szleifer and R. H. Bisseling, unpublished (1991).}
\bibitem{BMW94}
{\frenchspacing G. T. Barkema, J. F. Marko and B. Widom, Phys. Rev. E
{\bf49}, 5303 (1994).}
\bibitem{DEHP93}
{\frenchspacing B. Derrida, M. R. Evans, V. Hakim, and V. Pasquier,
J. Phys. A Math. Gen. {\bf 26}, 1493 (1993).}
\bibitem{SS95}
{\frenchspacing R. B. Stinchcombe and G. M. Sch\"utz, Phys. Rev. Lett.
{\bf 75}, 140 (1995).}
\bibitem{Note1}
It can be shown that the rank $R_N$ of the matrix for a system of $N$
reptons obeys the relation $R_N = 4(R_{N-1} - R_{N-2})$, and given this
result the expression for $R_N$ above follows by induction.
\bibitem{PS96}
{\frenchspacing M. Pr\"ahofer and H. Spohn, Physica A {\bf233}, 191 (1996).}
\end{references}
\end{document}